\documentclass[aps,prl,twocolumn,showpacs,superscriptaddress,groupedaddress,preprintnumbers,floatfix,reprint,superscriptaddress]{revtex4-1}

\usepackage{graphicx}  
\usepackage{dcolumn}   
\usepackage{bm}        
\usepackage{amssymb}   

\usepackage{subfigure}
\usepackage{multirow}
\usepackage{placeins}
\usepackage{morefloats}
\usepackage{epstopdf}

\usepackage{amsfonts}    
\usepackage{amsmath}
\usepackage{mathtools}

\def\GeV{\;\! \mathrm{GeV}}
\def\fm{\;\! \mathrm{fm}}

\allowdisplaybreaks

\begin{document}

\preprint{
\vbox{
\hbox{ADP-15-8/T910}
\hbox{Edinburgh 2015/02}
\hbox{LTH 1037}
\hbox{DESY 15-022}
}}

\title{Charge Symmetry Violation in the Electromagnetic Form Factors of the Proton}

\author{P.E.~Shanahan}\affiliation{ARC Centre of Excellence in Particle Physics at the Terascale and CSSM, School of Physical Sciences, University of Adelaide, Adelaide SA 5005, Australia}
\author{R.~Horsley}\affiliation{School of Physics and Astronomy, University of Edinburgh, Edinburgh EH9 3FD, UK}
\author{Y.~Nakamura}\affiliation{RIKEN Advanced Institute for Computational Science, Kobe, Hyogo 650-0047, Japan}
\author{D.~Pleiter}\affiliation{JSC, Forschungzentrum J\"ulich, 52425 J\"ulich, Germany} \affiliation{Institut f\"ur Theoretische Physik, Universit\"at Regensburg, 93040 Regensburg, Germany}
\author{P.E.L.~Rakow}\affiliation{Theoretical Physics Division, Department of Mathematical Sciences, University of Liverpool, Liverpool L69 3BX, UK}
\author{G.~Schierholz}\affiliation{Deutsches Elektronen-Synchrotron DESY, 22603 Hamburg, Germany}
\author{H.~St\"uben}\affiliation{Regionales Rechenzentrum, Universit\"at Hamburg, 20146 Hamburg, Germany}
\author{A.W.~Thomas}\affiliation{ARC Centre of Excellence in Particle Physics at the Terascale and CSSM, School of Physical Sciences, University of Adelaide, Adelaide SA 5005, Australia}
\author{R.D.~Young}\affiliation{ARC Centre of Excellence in Particle Physics at the Terascale and CSSM, School of Physical Sciences, University of Adelaide, Adelaide SA 5005, Australia}
\author{J.M.~Zanotti}\affiliation{ARC Centre of Excellence in Particle Physics at the Terascale and CSSM, School of Physical Sciences, University of Adelaide, Adelaide SA 5005, Australia}

\collaboration{CSSM and QCDSF/UKQCD Collaborations}

\begin{abstract}
Experimental tests of QCD through its predictions for the strange-quark content of the proton have been drastically restricted by our lack of knowledge of the violation of charge symmetry (CSV). We find unexpectedly tiny CSV in the proton's electromagnetic form factors by performing the first extraction of these quantities based on an analysis of lattice QCD data. 
The resulting values are an order of magnitude smaller than current bounds on proton strangeness from parity violating electron-proton scattering experiments. 
This result paves the way for a new generation of experimental measurements of the proton's strange form factors to challenge the predictions of QCD.

\end{abstract}


\maketitle

Charge symmetry is the invariance of the strong interaction under an isospin rotation exchanging $u$ and $d$ quarks (i.e., exchanging the proton and neutron). The violation of this symmetry (CSV) is arguably small: the proton-neutron mass difference is one part in a thousand~\cite{Borsanyi:2014jba} and many nuclear reactions proceed identically if protons and neutrons are interchanged. The effects of this small CSV, however, may be hugely significant. For example, if the proton-neutron mass difference were reversed protons could decay and atoms could not form. Charge symmetry violation also explains the discrepancy between the calculated and measured binding energy differences of mirror nuclei (Okamoto-Nolen-Schiffer anomaly)~\cite{Nolen:1969ms,Negele1971} and may play a role in precision tests of the Standard Model~\cite{Horsley:2010th}, including those at the LHC~\cite{Londergan:2009kj}.

In the late 1980s it was suggested that one could use measurements of neutral weak current matrix elements by parity-violating electron scattering (PVES)~\cite{Kaplan:1988ku,Mckeown:1989ir,Beck:1989tg} to determine the contribution of strange quark-antiquark pairs to the elastic electroweak form factors of the nucleon. These `strange form factors' have been the focus of intensive experimental and theoretical effort for the past two decades~\cite{Armstrong:2012bi}.
At present, the accuracy of theoretical calculations of these quantities~\cite{Shanahan:2014tja,Doi:2009sq,Leinweber:2006ug,Leinweber:2004tc} exceeds that of the best experimental values~\cite{Young:2007zs} by almost an order of magnitude---a remarkable exception in strong-interaction physics.
The limiting factor in future state-of-the-art PVES measurements at Mainz~\cite{Maas:2004dh,Baunack:2009gy} and JLab~\cite{Aniol:2005zg,Aniol:2005zf,Acha:2006my} is theoretical, arising from the assumption that CSV in the proton's electromagnetic form factors is negligible. 

Precisely, CSV form factors $G_\textrm{CSV}$, if not accounted for, mimic the strange-quark contribution $G^s_{E/M}$ in the combination of form factors accessed by experiment: the measured neutral weak current matrix elements $G^{p,Z}_{E/M}$ may be expressed as  
\begin{equation}\label{eq:FFeq}
G^{p,Z}_{E/M} = \left(1-4\sin^2\theta_W \right)G^{p,\gamma}_{E/M} - G^{n,\gamma}_{E/M} - G^s_{E/M} + G_\textrm{CSV},
\end{equation}
where the weak mixing-angle, $\theta_W$, and the total electromagnetic form factors, $G^{p/n,\gamma}_{E/M}$, are precisely determined from other experimental studies. 
With theoretical predictions of the size of $G_\textrm{CSV}$ varying through several orders of magnitude~\cite{Wagman:2014nfa,Kubis:2006cy,Miller:2006tv}, this uncertainty has halted experimental parity-violating electron scattering programs~\cite{Acha:2006my}.

In this Letter we report the first determination of CSV in the proton's electromagnetic form factors based on an analysis of lattice QCD data.
In terms of individual $u$ and $d$-quark contributions to the Sachs electric and magnetic form factors of the proton and neutron (conventionally defined without the charge factors), the CSV form factors which we calculate are defined as
\begin{equation} \label{eq:du}
\delta^u_{E/M} = G_{E/M}^{p,u}-G_{E/M}^{n,d},
\hspace{5mm}
\delta^d_{E/M} = G_{E/M}^{p,d}-G_{E/M}^{n,u},
\end{equation}
where we explicitly calculate $G_{E/M}^{p/n,u/d}$ and perform the subtractions indicated.
The combination relevant to experimental determinations of nucleon strangeness using Eq.~\eqref{eq:FFeq} is
\begin{equation}\label{eq:Gcsv}
G_\textrm{CSV} =  \left( \frac{2}{3}\delta^d_{E/M} - \frac{1}{3}\delta^u_{E/M}\right).
\end{equation}

The lattice results used here are an extension of those reported in Refs.~\cite{Shanahan:2014Elec,Shanahan:2014uka}; we include two independent sets of $2+1$-flavor simulations at different values of the finite lattice spacing $a$. 
Any discretization artifacts should appear at $\mathcal{O}(a^2)$. 
Each set consists of results for the individual connected quark contributions to the electromagnetic form factors of the entire outer-ring baryon octet at a range of pion masses down to 220~MeV and at 6 (set I) or 7 (set II) fixed values of the momentum transfer $Q^2$ up to 1.4~GeV$^2$. These values of $Q^2$ are relevant to experimental studies of the strange nucleon form factors. The lattice volumes are $L^3\times T=32^3 \times 64$ and $48^3 \times 96$, and the lattice spacings are $a^2=0.0055(3)$~fm$^2$ and $0.0038(2)$~fm$^2$ (set using various singlet quantities~\cite{Horsley:2013wqa,Bietenholz:2011qq}) for the two sets respectively.

Our extraction of the CSV form factors from the lattice simulations is based on the extrapolation of those results to infinite volume and to the physical pseudoscalar masses using a formalism based on connected chiral perturbation theory~\cite{Leinweber:2002qb,Tiburzi:2009yd}.
The extrapolation procedure is detailed in Refs.~\cite{Shanahan:2014Elec,Shanahan:2014uka}. 
The small finite-volume corrections are model-independent and the chiral extrapolation is demonstrated to be under control---the fit includes lattice data at low meson masses within the convergence regime of the effective theory, and it reproduces the experimental form factors at the physical masses~\cite{Shanahan:2014Elec,Shanahan:2014uka}.
To determine the CSV terms we must extend that work to incorporate the breaking of the flavor-SU(2) symmetry, i.e., to allow for unequal light quark masses, $m_u\ne m_d$.
This is a simple extension, and is performed precisely as in previous work where the same procedure was used to evaluate the mass splittings among members of baryon isospin multiplets~\cite{Shanahan:2012wa}, the CSV sigma terms~\cite{Shanahan:2013cd}, and the CSV parton distribution moments~\cite{Shanahan:2013vla} from 2+1-flavor lattice simulation results.
In brief, the low-energy parameters which appear in the SU(2)-breaking terms in the chiral extrapolation expressions for the CSV form factors also appear in the isospin-averaged expressions. These parameters are thus fixed by the fits to the $N_f=2+1$ lattice QCD simulations on the baryon octet which are presented in Refs.~\cite{Shanahan:2014Elec,Shanahan:2014uka}.

In principle, the CSV form factors on an infinite volume and at the physical pseudoscalar masses may thus, given the extrapolations of Refs.~\cite{Shanahan:2014Elec,Shanahan:2014uka}, be evaluated simply by performing the subtractions shown in Eq.~\eqref{eq:du}. 
This procedure, however, suffers from a significant systematic effect resulting from the omission of quark-line disconnected contributions in the simulations. To account for this omission we use the chiral extrapolation expressions to model the disconnected pieces of the loop integral expressions. This amounts to the replacement of the `connected' extrapolation coefficients of Refs.~\cite{Shanahan:2014Elec,Shanahan:2014uka} by the `full' expressions, where the free parameters remain as fixed by the connected fits.
The resulting expressions for the CSV electric and magnetic form factors (including disconnected quark-line contributions) as a function of meson masses can be written as 
\begin{widetext}
\begin{align} \nonumber
\delta^u_M= &\frac{1}{6}\left(2 c^M_1 - 3 c^M_{10} - 3 c^M_{12} - 4 c^M_2 - 2 c^M_5 - 5 c^M_6 - 54 c^M_7 + 3 c^M_9\right)\mathcal{B}(m_d-m_u) \\ 
& + \frac{M_N}{16\pi^3f_\pi^2}\frac{1}{9} \left[\mathcal{C}^2 \left(I_D^M(m_{K^0}) - I_D^M(m_{K^\pm})\right)- 12 \left(D^2 + 3 F^2\right) \left(I_O^M(m_{K^0}) - I_O^M(m_{K^\pm})\right)\right],\\[5pt] \nonumber
\delta^d_M= &\frac{1}{6} \left(2 c^M_1 + 2 c^M_{10} - 4 c^M_{11} + 2 c^M_{12} - 4 c^M_2 + 4 c^M_5 + c^M_6 + 54 c^M_7 - c^M_9\right)\mathcal{B}(m_d-m_u) \\\label{eq:CSV1}
& - \frac{M_N}{16\pi^3f_\pi^2}\frac{2}{9} \left[\mathcal{C}^2 \left(I_D^M(m_{K^0}) - I_D^M(m_{K^\pm})\right) - 9 \left(D - F\right)^2 \left(I_O^M(m_{K^0}) - I_O^M(m_{K^\pm})\right)\right],\\[10pt]
\nonumber
\delta^u_E= &\frac{1}{6}\left(2 c^E_1 - 3 c^E_{10} - 3 c^E_{12} - 4 c^E_2 - 2 c^E_5 - 5 c^E_6 - 54 c^E_7 + 3 c^E_9\right)Q^2\mathcal{B}(m_d-m_u) \\\nonumber
& - \frac{1}{16\pi^3f_\pi^2}\frac{1}{9} \left[\mathcal{C}^2 \left(I_D^E(m_{K^0}) - I_D^E(m_{K^\pm})\right) + 6 \left(D^2 + 3 F^2\right) \left(I_O^E(m_{K^0}) - I_O^E(m_{K^\pm})\right)\right.\\
& \hspace{2cm} \left.+18\left(I_T^E(m_{K^0}) - I_T^E(m_{K^\pm})\right)\right],\\[5pt] \nonumber
\delta^d_E= &\frac{1}{6} \left(2 c^E_1 + 2 c^E_{10} - 4 c^E_{11} + 2 c^E_{12} - 4 c^E_2 + 4 c^E_5 + c^E_6 + 54 c^E_7 - c^E_9\right)Q^2\mathcal{B}(m_d-m_u) \\ \nonumber
&+ \frac{1}{16\pi^3f_\pi^2}\frac{1}{9} \left[2\mathcal{C}^2 \left(I_D^E(m_{K^0}) - I_D^E(m_{K^\pm})\right) + 9 \left(D - F\right)^2 \left(I_O^E(m_{K^0}) - I_O^E(m_{K^\pm})\right)\right. \\ \label{eq:CSV2}
& \hspace{2.3cm}\left.+9\left(I_T^E(m_{K^0}) - I_T^E(m_{K^\pm})\right)\right],
\end{align}
\end{widetext}
where all symbols, including the low-energy constants $c_i^{E/M}$, are defined in Refs.~\cite{Shanahan:2014Elec,Shanahan:2014uka}. The leading-order loop integral expressions include meson loops with octet-baryon ($I_O$) or decuplet-baryon ($I_D$) intermediate states, as well as tadpole loops ($I_T$). The Gell-Mann-Oakes-Renner relation suggests the definition
\begin{equation}
\mathcal{B}(m_d-m_u) = \frac{(1-R)}{(1+R)}m_\pi^2,
\end{equation}
where $R$ denotes the light-quark mass ratio $R=m_u/m_d$. We take $R=0.553(43)$, determined by a fit to meson decay rates~\cite{Leutwyler:1996qg}. The final results are all consistent within uncertainties if we instead take the FLAG value $R=0.46(2)(2)$~\cite{FLAG}.

\begin{figure}
\centering
\subfigure[b][]{
\includegraphics[width=0.16\textwidth]{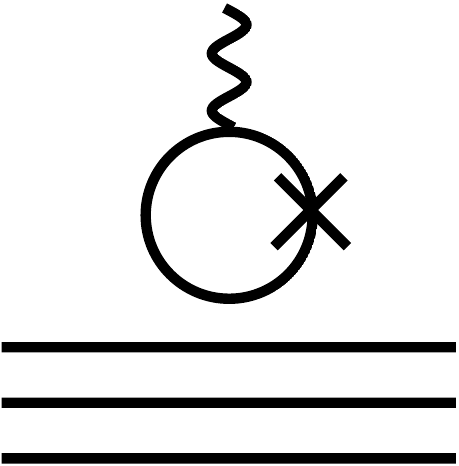}
\label{fig:insc72}
}\quad
\subfigure[b][]{
\includegraphics[width=0.16\textwidth]{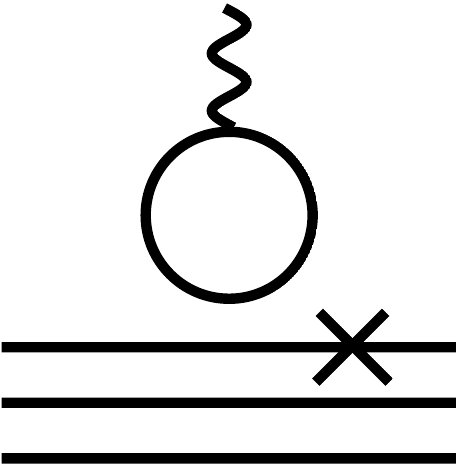}
\label{fig:insc122}
}
\caption{{Diagrammatic quark-line skeleton representation of omitted contributions to the CSV form factors.} Solid and wavy lines represent quarks and photons respectively. The crosses denote quark mass insertions, i.e., the figures represent the contribution from disconnected quark loops to CSV arising from the different ($u$ and $d$ quark) masses of: (a): the struck sea quark; (b): spectator quarks. These contributions are proportional to $\mathcal{B}(m_d-m_u)$.} \label{fig:diags}
\end{figure}

All of the low-energy parameters, other than $c_1^{E/M}$, $c_2^{E/M}$ and $c_7^{E/M}$, are determined from the chiral fits to the connected contribution to the isospin-averaged electromagnetic form factors which are described in Refs.~\cite{Shanahan:2014Elec,Shanahan:2014uka}.
While this procedure systematically includes some of the disconnected contribution to the CSV form factors, other disconnected terms---those which are linear in $\mathcal{B}(m_d-m_u)$ and not generated by chiral logarithms from meson loops---cannot be determined in this way.
Precisely, the terms which are generated by the Lagrangian pieces with coefficients $c_1^{E/M}$, $c_2^{E/M}$ and $c_7^{E/M}$ cannot be determined from the present lattice simulations. Physically, these terms arise from the diagrams illustrated and described in Fig.~\ref{fig:diags}. These contributions are anticipated to be small based on the success of valence quark models in reproducing form factor data. This is also supported by the results of direct lattice QCD calculations of $G_{E/M}$ which find that the disconnected contributions at small finite momentum transfer are consistent with zero and are bounded at the 1\% level~\cite{Abdel-Rehim:2013wlz}. The terms corresponding to the low-energy parameters $c_1^{E/M}$, $c_2^{E/M}$ and $c_7^{E/M}$ are only part of that small disconnected contribution.

\begin{figure}
\centering
\subfigure[b][]{
\includegraphics[width=0.18\textwidth]{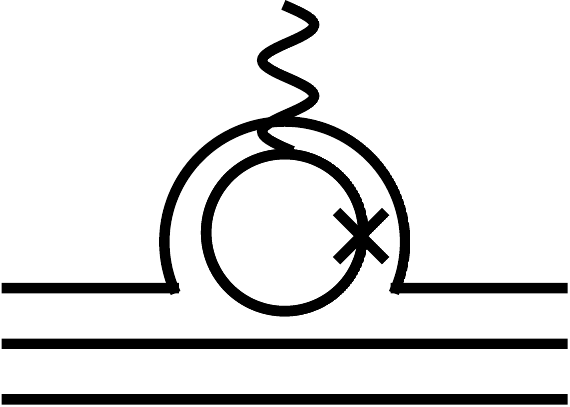}
\label{fig:LoopSub1}
}\quad
\subfigure[b][]{
\includegraphics[width=0.18\textwidth]{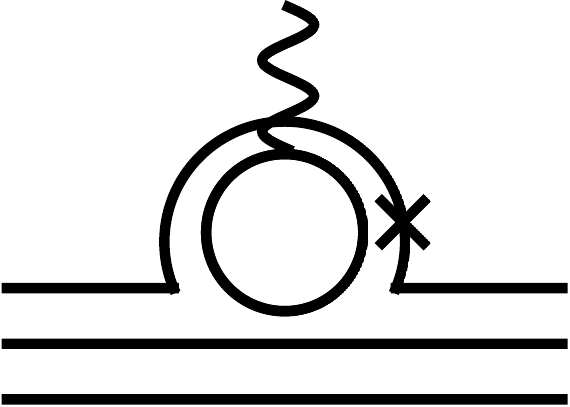}
\label{fig:LoopSub2}
}
\caption{\label{fig:subs}{Quark-line skeleton diagrams of the meson loops used to model the omitted contributions to the CSV form factors.} Solid and wavy lines represent quarks and photons respectively. The crosses denote quark mass insertions into: (a): the struck sea quark in the meson loop; (b): the meson loop spectator quark.} 
\end{figure}

We choose to set contributions from the unknown $c_1^{E/M}$, $c_2^{E/M}$ and $c_7^{E/M}$ terms to $0$, with an uncertainty taken to be twice the magnitude of the corresponding contributions from meson loop diagrams, evaluated with a dipole cutoff regulator with mass scale $\Lambda=0.8(2)\GeV$. We suggest that this error estimate is extremely conservative. The use of this method to evaluate the loops is justified by the well-established and successful use of this model to relate full and partially-quenched lattice QCD calculations~\cite{Wang:2014nhf}. 
The loop diagram used to estimate the $c^{E/M}_{1,2}$ terms is represented in Fig.~\ref{fig:LoopSub2}, where only the `loop spectator' quark mass (i.e., the valence-quark part of the meson mass) is changed. For the $c^{E/M}_7$ term, represented in Fig.~\ref{fig:LoopSub1}, only the sea-quark part of the loop meson mass is considered. These contributions are added in quadrature.
The magnitude of this contribution to the total uncertainty varies with $Q^2$; it is largest at our lowest $Q^2$-values where it contributes 20--60\% of the quoted uncertainty on the final result (depending which of $\delta_{E/M}^{u/d}$ one is considering), while at larger values of $Q^2$, consistent with the suppression of meson loops at high-$Q^2$, it contributes 1--15\%.

\begin{figure*}
\centering
\subfigure[l][]{
\includegraphics[width=0.475\textwidth]{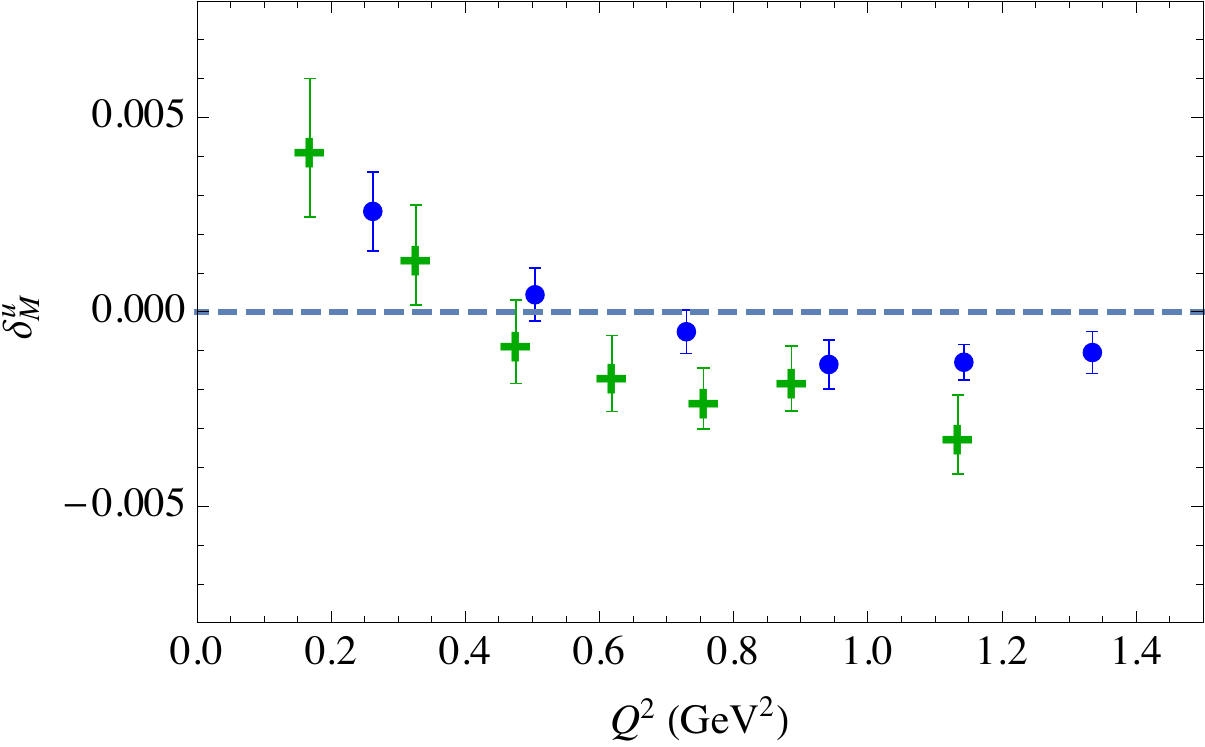}
}
\subfigure[r][]{
\includegraphics[width=0.475\textwidth]{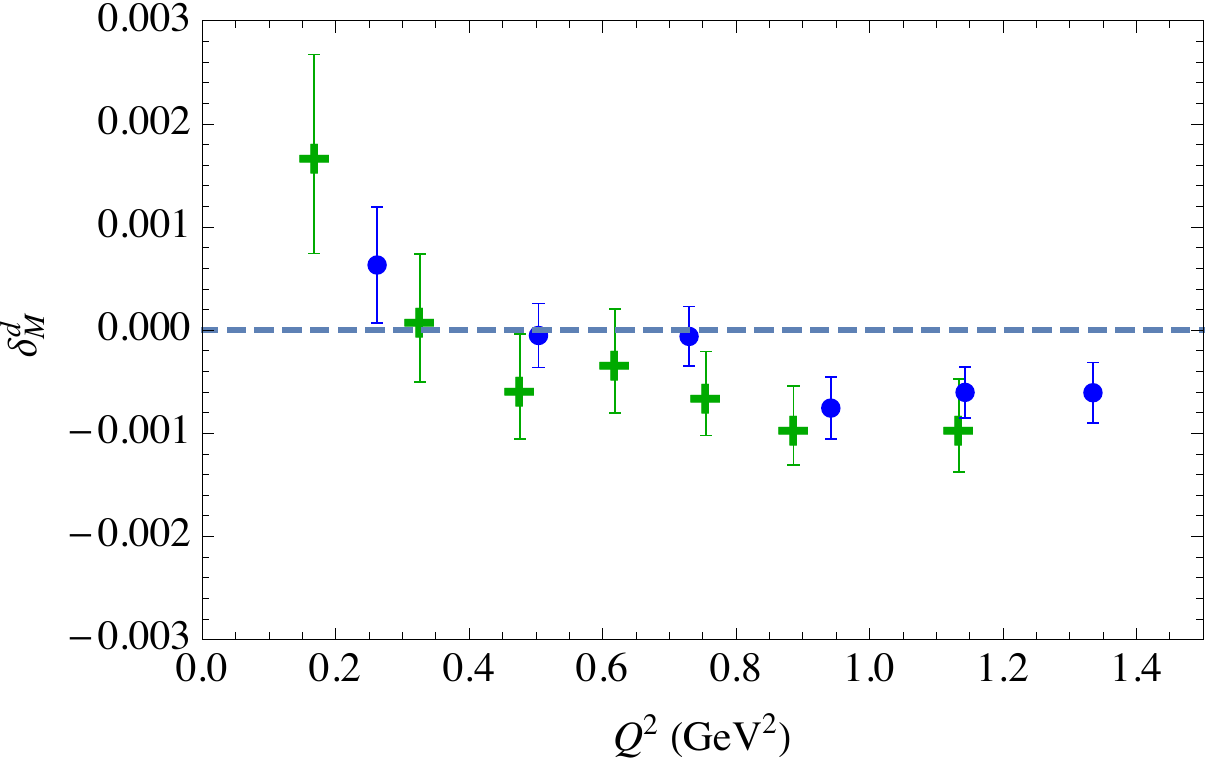}
}
\subfigure[l][]{
\includegraphics[width=0.475\textwidth]{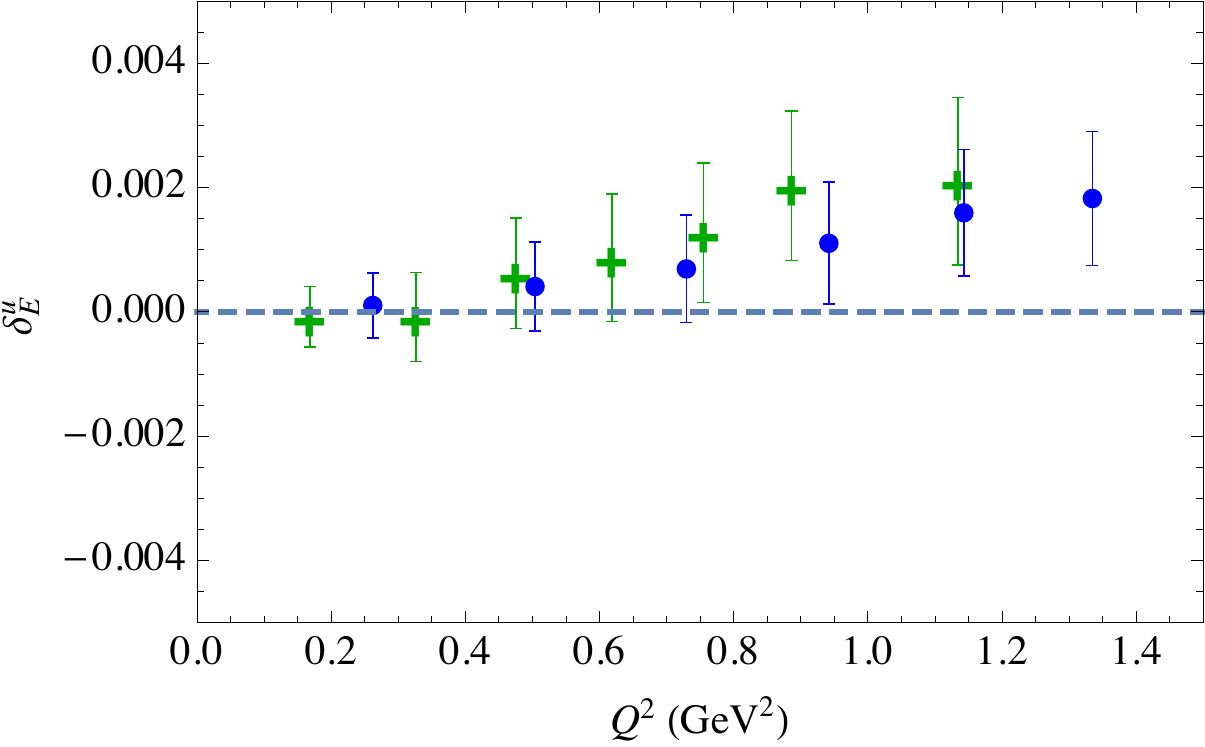}
}
\subfigure[r][]{
\includegraphics[width=0.475\textwidth]{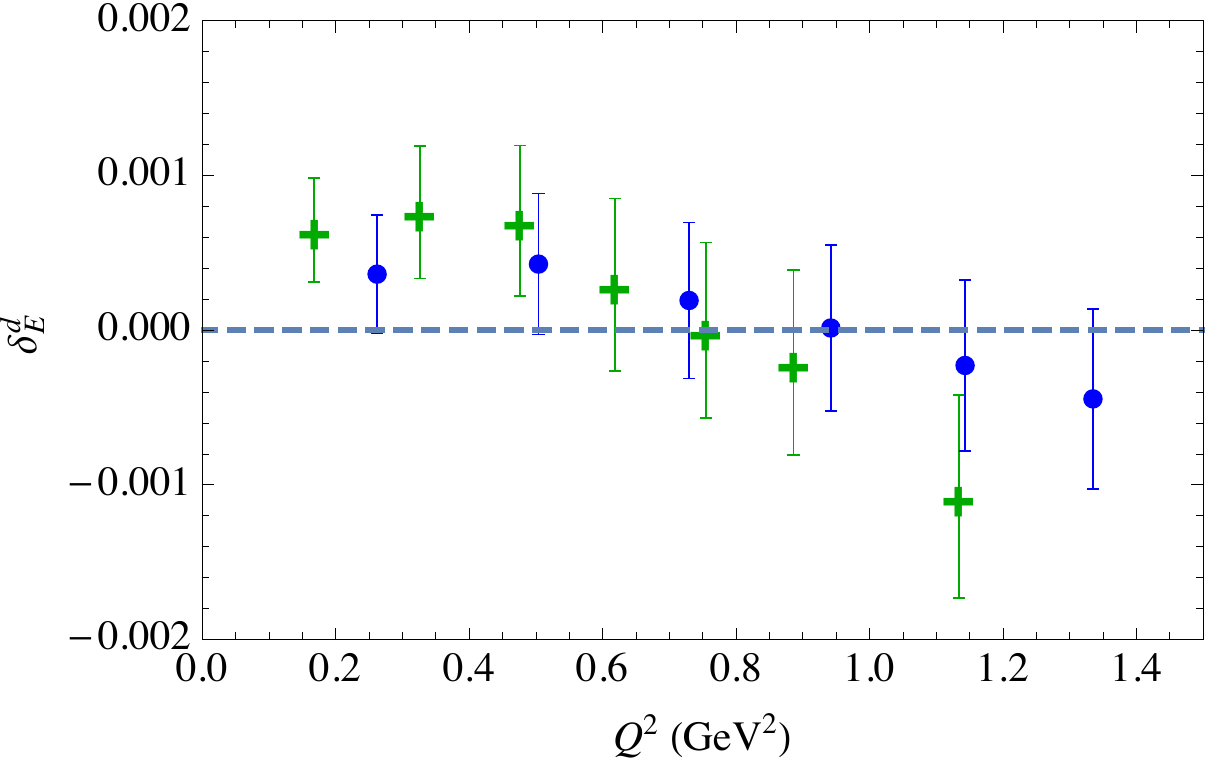}
}
\caption{\label{fig:individualCSV}Individual up and down quark contributions to the CSV form factors. These terms are combined to give the total CSV form factors $G_\textrm{CSV} =  \left( \frac{2}{3}\delta^d_{E/M} - \frac{1}{3}\delta^u_{E/M}\right)$. Blue points and green crosses show the results of data sets I and II extrapolated to the physical point, with corrections applied to model the omitted disconnected terms.}
\end{figure*}

\begin{figure*}
\centering
\subfigure[]{\label{fig:CSVFFM}
\includegraphics[width=0.475\textwidth]{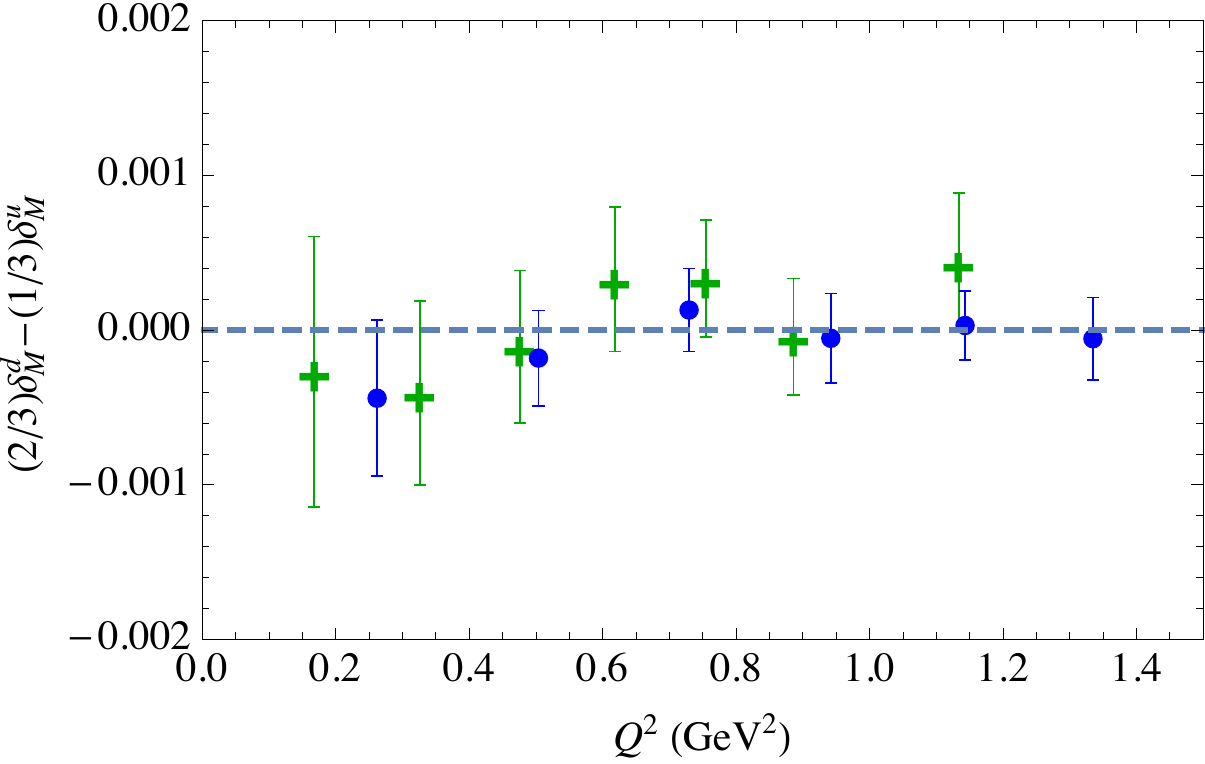}
}
\subfigure[]{\label{fig:CSVFFE}
\includegraphics[width=0.475\textwidth]{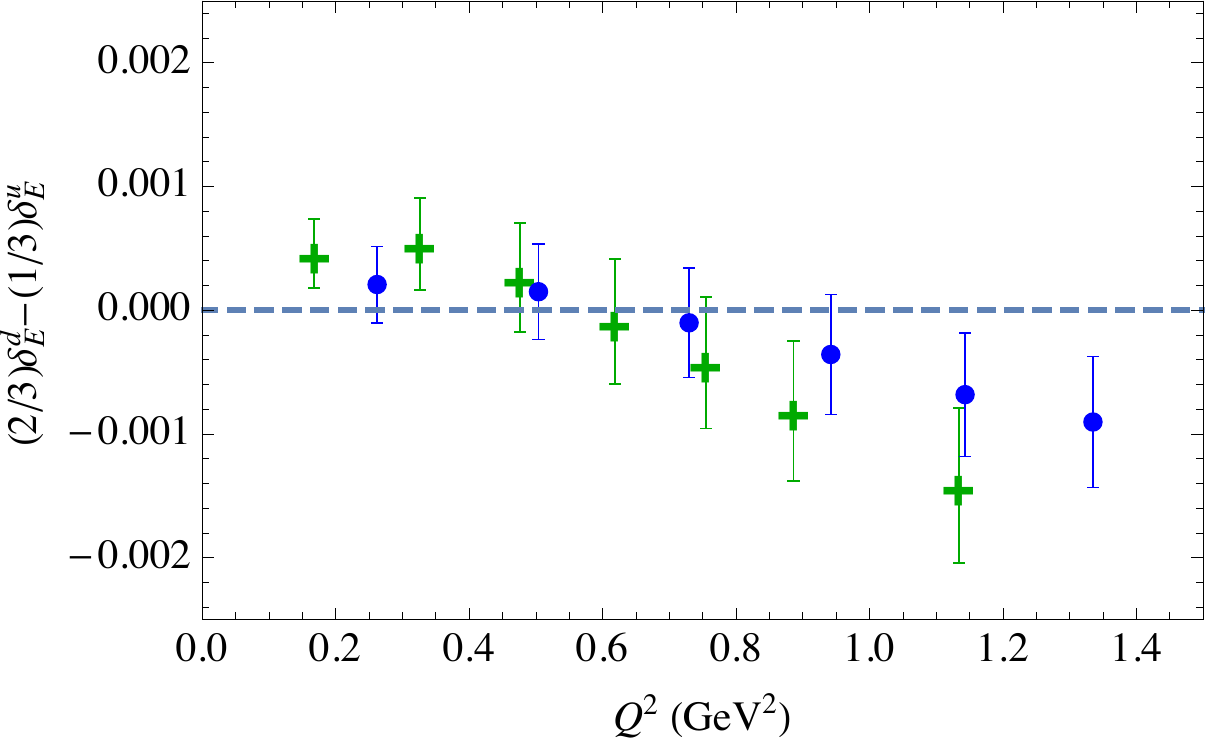}
}
\caption{\label{fig:TotalCSV}Magnetic and electric CSV form factors as relevant to experimental determinations of nucleon strangeness. The blue circles and green crosses 
denote our results based on simulation sets I ($a^2=0.0055(3)\fm^2$) and II ($a^2=0.0038(2)\fm^2$), respectively.}
\end{figure*}

The results of this analysis for the individual $u$ and $d$-quark contributions to the CSV electric and magnetic form factors of the proton are shown in Fig.~\ref{fig:individualCSV}. The close agreement of the two sets of simulations (at different lattice spacings $a$ and on different simulation volumes) confirms that the finite-volume corrections and chiral extrapolations are under control and that any discretization effects resulting from the finite lattice spacing are small.
The size of the CSV form factor combination, $G_\textrm{CSV}$, relevant to PVES experiments probing the strange electric and magnetic form factors of the nucleon by Eq.~\eqref{eq:FFeq}, is shown in Fig.~\ref{fig:TotalCSV}.
This result gives quantitative confirmation that CSV effects in the electromagnetic form factors, for momentum transfers up to approximately $1.4\GeV^2$, are at the level of 0.2\% of the relevant proton form factors---an order of magnitude smaller than the precision of existing PVES studies. 
To put this in perspective, the level of CSV shown in Fig.~\ref{fig:CSVFFE} is equivalent to a CSV difference in charge radii of less than one attometer.
These precise results open the door for a new generation of experiments to probe the structure of the quantum vacuum through the strange quark form factors.

\section*{Acknowledgements}
The numerical configuration generation was performed using the BQCD lattice QCD program~\cite{Nakamura:2010qh} on the IBM BlueGeneQ using DIRAC 2 resources (EPCC, Edinburgh, UK), the BlueGene P and Q at NIC (J\"ulich, Germany) and the Cray XC30 at HLRN (Berlin-Hannover, Germany). The BlueGene codes were optimised using Bagel~\cite{Boyle:2009vp}. The Chroma software library~\cite{Edwards:2004sx} was used in the data analysis. This work was supported by the EU grants 283286 (HadronPhysics3), 227431 (Hadron Physics2) and by the University of Adelaide and the Australian Research Council through the ARC Centre of Excellence for Particle Physics at the Terascale and grants FL0992247 (AWT), FT120100821 (RDY), DP140103067 (RDY and JMZ) and FT100100005 (JMZ).

\section*{References}
\bibliography{CSVBib}

\begin{thebibliography}{38}%
\makeatletter
\providecommand \@ifxundefined [1]{%
 \@ifx{#1\undefined}
}%
\providecommand \@ifnum [1]{%
 \ifnum #1\expandafter \@firstoftwo
 \else \expandafter \@secondoftwo
 \fi
}%
\providecommand \@ifx [1]{%
 \ifx #1\expandafter \@firstoftwo
 \else \expandafter \@secondoftwo
 \fi
}%
\providecommand \natexlab [1]{#1}%
\providecommand \enquote  [1]{``#1''}%
\providecommand \bibnamefont  [1]{#1}%
\providecommand \bibfnamefont [1]{#1}%
\providecommand \citenamefont [1]{#1}%
\providecommand \href@noop [0]{\@secondoftwo}%
\providecommand \href [0]{\begingroup \@sanitize@url \@href}%
\providecommand \@href[1]{\@@startlink{#1}\@@href}%
\providecommand \@@href[1]{\endgroup#1\@@endlink}%
\providecommand \@sanitize@url [0]{\catcode `\\12\catcode `\$12\catcode
  `\&12\catcode `\#12\catcode `\^12\catcode `\_12\catcode `\%12\relax}%
\providecommand \@@startlink[1]{}%
\providecommand \@@endlink[0]{}%
\providecommand \url  [0]{\begingroup\@sanitize@url \@url }%
\providecommand \@url [1]{\endgroup\@href {#1}{\urlprefix }}%
\providecommand \urlprefix  [0]{URL }%
\providecommand \Eprint [0]{\href }%
\providecommand \doibase [0]{http://dx.doi.org/}%
\providecommand \selectlanguage [0]{\@gobble}%
\providecommand \bibinfo  [0]{\@secondoftwo}%
\providecommand \bibfield  [0]{\@secondoftwo}%
\providecommand \translation [1]{[#1]}%
\providecommand \BibitemOpen [0]{}%
\providecommand \bibitemStop [0]{}%
\providecommand \bibitemNoStop [0]{.\EOS\space}%
\providecommand \EOS [0]{\spacefactor3000\relax}%
\providecommand \BibitemShut  [1]{\csname bibitem#1\endcsname}%
\let\auto@bib@innerbib\@empty
\bibitem [{\citenamefont {Borsanyi}\ \emph {et~al.}(2014)\citenamefont
  {Borsanyi}, \citenamefont {D\"urr}, \citenamefont {Fodor}, \citenamefont
  {Hoelbling}, \citenamefont {Katz} \emph {et~al.}}]{Borsanyi:2014jba}%
  \BibitemOpen
  \bibfield  {author} {\bibinfo {author} {\bibfnamefont {S.}~\bibnamefont
  {Borsanyi}}, \bibinfo {author} {\bibfnamefont {S.}~\bibnamefont {D\"urr}},
  \bibinfo {author} {\bibfnamefont {Z.}~\bibnamefont {Fodor}}, \bibinfo
  {author} {\bibfnamefont {C.}~\bibnamefont {Hoelbling}}, \bibinfo {author}
  {\bibfnamefont {S.}~\bibnamefont {Katz}},  \emph {et~al.} (\bibinfo
  {collaboration} {BMW Collaboration}),\ }\href@noop {} {\  (\bibinfo {year}
  {2014})},\ \Eprint {http://arxiv.org/abs/1406.4088} {arXiv:1406.4088
  [hep-lat]} \BibitemShut {NoStop}%
\bibitem [{\citenamefont {Nolen}\ and\ \citenamefont
  {Schiffer}(1969)}]{Nolen:1969ms}%
  \BibitemOpen
  \bibfield  {author} {\bibinfo {author} {\bibfnamefont {J.~A.}\ \bibnamefont
  {Nolen}}\ and\ \bibinfo {author} {\bibfnamefont {J.~P.}\ \bibnamefont
  {Schiffer}},\ }\href {\doibase 10.1146/annurev.ns.19.120169.002351}
  {\bibfield  {journal} {\bibinfo  {journal} {Ann. Rev. Nucl. Part. Sci.}\
  }\textbf {\bibinfo {volume} {19}},\ \bibinfo {pages} {471} (\bibinfo {year}
  {1969})}\BibitemShut {NoStop}%
\bibitem [{\citenamefont {Negele}(1971)}]{Negele1971}%
  \BibitemOpen
  \bibfield  {author} {\bibinfo {author} {\bibfnamefont {J.~W.}\ \bibnamefont
  {Negele}},\ }\href@noop {} {\bibfield  {journal} {\bibinfo  {journal} {Nucl.
  Phys.}\ }\textbf {\bibinfo {volume} {A165}},\ \bibinfo {pages} {305}
  (\bibinfo {year} {1971})}\BibitemShut {NoStop}%
\bibitem [{\citenamefont {Horsley}\ \emph {et~al.}(2011)\citenamefont
  {Horsley}, \citenamefont {Nakamura}, \citenamefont {Pleiter}, \citenamefont
  {Rakow}, \citenamefont {Schierholz} \emph {et~al.}}]{Horsley:2010th}%
  \BibitemOpen
  \bibfield  {author} {\bibinfo {author} {\bibfnamefont {R.}~\bibnamefont
  {Horsley}}, \bibinfo {author} {\bibfnamefont {Y.}~\bibnamefont {Nakamura}},
  \bibinfo {author} {\bibfnamefont {D.}~\bibnamefont {Pleiter}}, \bibinfo
  {author} {\bibfnamefont {P.}~\bibnamefont {Rakow}}, \bibinfo {author}
  {\bibfnamefont {G.}~\bibnamefont {Schierholz}},  \emph {et~al.},\ }\href
  {\doibase 10.1103/PhysRevD.83.051501} {\bibfield  {journal} {\bibinfo
  {journal} {Phys.Rev.}\ }\textbf {\bibinfo {volume} {D83}},\ \bibinfo {pages}
  {051501} (\bibinfo {year} {2011})},\ \Eprint {http://arxiv.org/abs/1012.0215}
  {arXiv:1012.0215 [hep-lat]} \BibitemShut {NoStop}%
\bibitem [{\citenamefont {Londergan}\ \emph {et~al.}(2010)\citenamefont
  {Londergan}, \citenamefont {Peng},\ and\ \citenamefont
  {Thomas}}]{Londergan:2009kj}%
  \BibitemOpen
  \bibfield  {author} {\bibinfo {author} {\bibfnamefont {J.~T.}\ \bibnamefont
  {Londergan}}, \bibinfo {author} {\bibfnamefont {J.~C.}\ \bibnamefont {Peng}},
  \ and\ \bibinfo {author} {\bibfnamefont {A.~W.}\ \bibnamefont {Thomas}},\
  }\href {\doibase 10.1103/RevModPhys.82.2009} {\bibfield  {journal} {\bibinfo
  {journal} {Rev. Mod. Phys.}\ }\textbf {\bibinfo {volume} {82}},\ \bibinfo
  {pages} {2009} (\bibinfo {year} {2010})},\ \Eprint
  {http://arxiv.org/abs/0907.2352} {arXiv:0907.2352 [hep-ph]} \BibitemShut
  {NoStop}%
\bibitem [{\citenamefont {Kaplan}\ and\ \citenamefont
  {Manohar}(1988)}]{Kaplan:1988ku}%
  \BibitemOpen
  \bibfield  {author} {\bibinfo {author} {\bibfnamefont {D.~B.}\ \bibnamefont
  {Kaplan}}\ and\ \bibinfo {author} {\bibfnamefont {A.}~\bibnamefont
  {Manohar}},\ }\href {\doibase 10.1016/0550-3213(88)90090-9} {\bibfield
  {journal} {\bibinfo  {journal} {Nucl. Phys.}\ }\textbf {\bibinfo {volume}
  {B310}},\ \bibinfo {pages} {527} (\bibinfo {year} {1988})}\BibitemShut
  {NoStop}%
\bibitem [{\citenamefont {Mckeown}(1989)}]{Mckeown:1989ir}%
  \BibitemOpen
  \bibfield  {author} {\bibinfo {author} {\bibfnamefont {R.~D.}\ \bibnamefont
  {Mckeown}},\ }\href {\doibase 10.1016/0370-2693(89)90364-X} {\bibfield
  {journal} {\bibinfo  {journal} {Phys. Lett.}\ }\textbf {\bibinfo {volume}
  {B219}},\ \bibinfo {pages} {140} (\bibinfo {year} {1989})}\BibitemShut
  {NoStop}%
\bibitem [{\citenamefont {Beck}(1989)}]{Beck:1989tg}%
  \BibitemOpen
  \bibfield  {author} {\bibinfo {author} {\bibfnamefont {D.~H.}\ \bibnamefont
  {Beck}},\ }\href {\doibase 10.1103/PhysRevD.39.3248} {\bibfield  {journal}
  {\bibinfo  {journal} {Phys. Rev.}\ }\textbf {\bibinfo {volume} {D39}},\
  \bibinfo {pages} {3248} (\bibinfo {year} {1989})}\BibitemShut {NoStop}%
\bibitem [{\citenamefont {Armstrong}\ and\ \citenamefont
  {McKeown}(2012)}]{Armstrong:2012bi}%
  \BibitemOpen
  \bibfield  {author} {\bibinfo {author} {\bibfnamefont {D.~S.}\ \bibnamefont
  {Armstrong}}\ and\ \bibinfo {author} {\bibfnamefont {R.~D.}\ \bibnamefont
  {McKeown}},\ }\href {\doibase 10.1146/annurev-nucl-102010-130419} {\bibfield
  {journal} {\bibinfo  {journal} {Ann. Rev. Nucl. Part. Sci.}\ }\textbf
  {\bibinfo {volume} {62}},\ \bibinfo {pages} {337} (\bibinfo {year} {2012})},\
  \Eprint {http://arxiv.org/abs/1207.5238} {arXiv:1207.5238 [nucl-ex]}
  \BibitemShut {NoStop}%
\bibitem [{\citenamefont {Shanahan}\ \emph {et~al.}(2015)\citenamefont
  {Shanahan}, \citenamefont {Horsley}, \citenamefont {Nakamura}, \citenamefont
  {Pleiter}, \citenamefont {Rakow} \emph {et~al.}}]{Shanahan:2014tja}%
  \BibitemOpen
  \bibfield  {author} {\bibinfo {author} {\bibfnamefont {P.}~\bibnamefont
  {Shanahan}}, \bibinfo {author} {\bibfnamefont {R.}~\bibnamefont {Horsley}},
  \bibinfo {author} {\bibfnamefont {Y.}~\bibnamefont {Nakamura}}, \bibinfo
  {author} {\bibfnamefont {D.}~\bibnamefont {Pleiter}}, \bibinfo {author}
  {\bibfnamefont {P.}~\bibnamefont {Rakow}},  \emph {et~al.},\ }\href@noop {}
  {\bibfield  {journal} {\bibinfo  {journal} {Phys. Rev. Lett.}\ }\textbf
  {\bibinfo {volume} {114}},\ \bibinfo {pages} {091802} (\bibinfo {year}
  {2015})},\ \Eprint {http://arxiv.org/abs/1403.6537} {arXiv:1403.6537
  [hep-lat]} \BibitemShut {NoStop}%
\bibitem [{\citenamefont {Doi}\ \emph {et~al.}(2009)\citenamefont {Doi} \emph
  {et~al.}}]{Doi:2009sq}%
  \BibitemOpen
  \bibfield  {author} {\bibinfo {author} {\bibfnamefont {T.}~\bibnamefont
  {Doi}} \emph {et~al.},\ }\href {\doibase 10.1103/PhysRevD.80.094503}
  {\bibfield  {journal} {\bibinfo  {journal} {Phys. Rev.}\ }\textbf {\bibinfo
  {volume} {D80}},\ \bibinfo {pages} {094503} (\bibinfo {year} {2009})},\
  \Eprint {http://arxiv.org/abs/0903.3232} {arXiv:0903.3232 [hep-ph]}
  \BibitemShut {NoStop}%
\bibitem [{\citenamefont {Leinweber}\ \emph {et~al.}(2006)\citenamefont
  {Leinweber} \emph {et~al.}}]{Leinweber:2006ug}%
  \BibitemOpen
  \bibfield  {author} {\bibinfo {author} {\bibfnamefont {D.~B.}\ \bibnamefont
  {Leinweber}} \emph {et~al.},\ }\href {\doibase 10.1103/PhysRevLett.97.022001}
  {\bibfield  {journal} {\bibinfo  {journal} {Phys. Rev. Lett.}\ }\textbf
  {\bibinfo {volume} {97}},\ \bibinfo {pages} {022001} (\bibinfo {year}
  {2006})},\ \Eprint {http://arxiv.org/abs/hep-lat/0601025}
  {arXiv:hep-lat/0601025} \BibitemShut {NoStop}%
\bibitem [{\citenamefont {Leinweber}\ \emph {et~al.}(2005)\citenamefont
  {Leinweber} \emph {et~al.}}]{Leinweber:2004tc}%
  \BibitemOpen
  \bibfield  {author} {\bibinfo {author} {\bibfnamefont {D.~B.}\ \bibnamefont
  {Leinweber}} \emph {et~al.},\ }\href {\doibase 10.1103/PhysRevLett.94.212001}
  {\bibfield  {journal} {\bibinfo  {journal} {Phys. Rev. Lett.}\ }\textbf
  {\bibinfo {volume} {94}},\ \bibinfo {pages} {212001} (\bibinfo {year}
  {2005})},\ \Eprint {http://arxiv.org/abs/hep-lat/0406002}
  {arXiv:hep-lat/0406002} \BibitemShut {NoStop}%
\bibitem [{\citenamefont {Young}\ \emph {et~al.}(2007)\citenamefont {Young},
  \citenamefont {Carlini}, \citenamefont {Thomas},\ and\ \citenamefont
  {Roche}}]{Young:2007zs}%
  \BibitemOpen
  \bibfield  {author} {\bibinfo {author} {\bibfnamefont {R.~D.}\ \bibnamefont
  {Young}}, \bibinfo {author} {\bibfnamefont {R.~D.}\ \bibnamefont {Carlini}},
  \bibinfo {author} {\bibfnamefont {A.~W.}\ \bibnamefont {Thomas}}, \ and\
  \bibinfo {author} {\bibfnamefont {J.}~\bibnamefont {Roche}},\ }\href
  {\doibase 10.1103/PhysRevLett.99.122003} {\bibfield  {journal} {\bibinfo
  {journal} {Phys. Rev. Lett.}\ }\textbf {\bibinfo {volume} {99}},\ \bibinfo
  {pages} {122003} (\bibinfo {year} {2007})},\ \Eprint
  {http://arxiv.org/abs/0704.2618} {arXiv:0704.2618 [hep-ph]} \BibitemShut
  {NoStop}%
\bibitem [{\citenamefont {Maas}\ \emph {et~al.}(2005)\citenamefont {Maas} \emph
  {et~al.}}]{Maas:2004dh}%
  \BibitemOpen
  \bibfield  {author} {\bibinfo {author} {\bibfnamefont {F.~E.}\ \bibnamefont
  {Maas}} \emph {et~al.},\ }\href {\doibase 10.1103/PhysRevLett.94.152001}
  {\bibfield  {journal} {\bibinfo  {journal} {Phys. Rev. Lett.}\ }\textbf
  {\bibinfo {volume} {94}},\ \bibinfo {pages} {152001} (\bibinfo {year}
  {2005})},\ \Eprint {http://arxiv.org/abs/nucl-ex/0412030}
  {arXiv:nucl-ex/0412030} \BibitemShut {NoStop}%
\bibitem [{\citenamefont {Baunack}\ \emph {et~al.}(2009)\citenamefont
  {Baunack}, \citenamefont {Aulenbacher}, \citenamefont {Balaguer~Rios},
  \citenamefont {Capozza}, \citenamefont {Diefenbach} \emph
  {et~al.}}]{Baunack:2009gy}%
  \BibitemOpen
  \bibfield  {author} {\bibinfo {author} {\bibfnamefont {S.}~\bibnamefont
  {Baunack}}, \bibinfo {author} {\bibfnamefont {K.}~\bibnamefont
  {Aulenbacher}}, \bibinfo {author} {\bibfnamefont {D.}~\bibnamefont
  {Balaguer~Rios}}, \bibinfo {author} {\bibfnamefont {L.}~\bibnamefont
  {Capozza}}, \bibinfo {author} {\bibfnamefont {J.}~\bibnamefont {Diefenbach}},
   \emph {et~al.},\ }\href {\doibase 10.1103/PhysRevLett.102.151803} {\bibfield
   {journal} {\bibinfo  {journal} {Phys. Rev. Lett.}\ }\textbf {\bibinfo
  {volume} {102}},\ \bibinfo {pages} {151803} (\bibinfo {year} {2009})},\
  \Eprint {http://arxiv.org/abs/0903.2733} {arXiv:0903.2733 [nucl-ex]}
  \BibitemShut {NoStop}%
\bibitem [{\citenamefont {Aniol}\ \emph
  {et~al.}(2006{\natexlab{a}})\citenamefont {Aniol} \emph
  {et~al.}}]{Aniol:2005zg}%
  \BibitemOpen
  \bibfield  {author} {\bibinfo {author} {\bibfnamefont {K.~A.}\ \bibnamefont
  {Aniol}} \emph {et~al.} (\bibinfo {collaboration} {HAPPEX Collaboration}),\
  }\href {\doibase 10.1016/j.physletb.2006.03.011} {\bibfield  {journal}
  {\bibinfo  {journal} {Phys. Lett.}\ }\textbf {\bibinfo {volume} {B635}},\
  \bibinfo {pages} {275} (\bibinfo {year} {2006}{\natexlab{a}})},\ \Eprint
  {http://arxiv.org/abs/nucl-ex/0506011} {arXiv:nucl-ex/0506011} \BibitemShut
  {NoStop}%
\bibitem [{\citenamefont {Aniol}\ \emph
  {et~al.}(2006{\natexlab{b}})\citenamefont {Aniol} \emph
  {et~al.}}]{Aniol:2005zf}%
  \BibitemOpen
  \bibfield  {author} {\bibinfo {author} {\bibfnamefont {K.~A.}\ \bibnamefont
  {Aniol}} \emph {et~al.} (\bibinfo {collaboration} {HAPPEX Collaboration}),\
  }\href {\doibase 10.1103/PhysRevLett.96.022003} {\bibfield  {journal}
  {\bibinfo  {journal} {Phys. Rev. Lett.}\ }\textbf {\bibinfo {volume} {96}},\
  \bibinfo {pages} {022003} (\bibinfo {year} {2006}{\natexlab{b}})},\ \Eprint
  {http://arxiv.org/abs/nucl-ex/0506010} {arXiv:nucl-ex/0506010} \BibitemShut
  {NoStop}%
\bibitem [{\citenamefont {Acha}\ \emph {et~al.}(2007)\citenamefont {Acha} \emph
  {et~al.}}]{Acha:2006my}%
  \BibitemOpen
  \bibfield  {author} {\bibinfo {author} {\bibfnamefont {A.}~\bibnamefont
  {Acha}} \emph {et~al.} (\bibinfo {collaboration} {HAPPEX Collaboration}),\
  }\href {\doibase 10.1103/PhysRevLett.98.032301} {\bibfield  {journal}
  {\bibinfo  {journal} {Phys. Rev. Lett.}\ }\textbf {\bibinfo {volume} {98}},\
  \bibinfo {pages} {032301} (\bibinfo {year} {2007})},\ \Eprint
  {http://arxiv.org/abs/nucl-ex/0609002} {arXiv:nucl-ex/0609002} \BibitemShut
  {NoStop}%
\bibitem [{\citenamefont {Wagman}\ and\ \citenamefont
  {Miller}(2014)}]{Wagman:2014nfa}%
  \BibitemOpen
  \bibfield  {author} {\bibinfo {author} {\bibfnamefont {M.}~\bibnamefont
  {Wagman}}\ and\ \bibinfo {author} {\bibfnamefont {G.~A.}\ \bibnamefont
  {Miller}},\ }\href {\doibase 10.1103/PhysRevC.89.065206} {\bibfield
  {journal} {\bibinfo  {journal} {Phys. Rev.}\ }\textbf {\bibinfo {volume}
  {C89}},\ \bibinfo {pages} {065206} (\bibinfo {year} {2014})},\ \Eprint
  {http://arxiv.org/abs/1402.7169} {arXiv:1402.7169 [nucl-th]} \BibitemShut
  {NoStop}%
\bibitem [{\citenamefont {Kubis}\ and\ \citenamefont
  {Lewis}(2006)}]{Kubis:2006cy}%
  \BibitemOpen
  \bibfield  {author} {\bibinfo {author} {\bibfnamefont {B.}~\bibnamefont
  {Kubis}}\ and\ \bibinfo {author} {\bibfnamefont {R.}~\bibnamefont {Lewis}},\
  }\href {\doibase 10.1103/PhysRevC.74.015204} {\bibfield  {journal} {\bibinfo
  {journal} {Phys. Rev.}\ }\textbf {\bibinfo {volume} {C74}},\ \bibinfo {pages}
  {015204} (\bibinfo {year} {2006})},\ \Eprint
  {http://arxiv.org/abs/nucl-th/0605006} {arXiv:nucl-th/0605006} \BibitemShut
  {NoStop}%
\bibitem [{\citenamefont {Miller}\ \emph {et~al.}(2006)\citenamefont {Miller},
  \citenamefont {Opper},\ and\ \citenamefont {Stephenson}}]{Miller:2006tv}%
  \BibitemOpen
  \bibfield  {author} {\bibinfo {author} {\bibfnamefont {G.~A.}\ \bibnamefont
  {Miller}}, \bibinfo {author} {\bibfnamefont {A.~K.}\ \bibnamefont {Opper}}, \
  and\ \bibinfo {author} {\bibfnamefont {E.~J.}\ \bibnamefont {Stephenson}},\
  }\href {\doibase 10.1146/annurev.nucl.56.080805.140446} {\bibfield  {journal}
  {\bibinfo  {journal} {Ann. Rev. Nucl. Part. Sci.}\ }\textbf {\bibinfo
  {volume} {56}},\ \bibinfo {pages} {253} (\bibinfo {year} {2006})},\ \Eprint
  {http://arxiv.org/abs/nucl-ex/0602021} {arXiv:nucl-ex/0602021} \BibitemShut
  {NoStop}%
\bibitem [{\citenamefont {Shanahan}\ \emph
  {et~al.}(2014{\natexlab{a}})\citenamefont {Shanahan} \emph
  {et~al.}}]{Shanahan:2014Elec}%
  \BibitemOpen
  \bibfield  {author} {\bibinfo {author} {\bibfnamefont {P.~E.}\ \bibnamefont
  {Shanahan}} \emph {et~al.},\ }\href@noop {} {\bibfield  {journal} {\bibinfo
  {journal} {Phys. Rev.}\ }\textbf {\bibinfo {volume} {D90}},\ \bibinfo {pages}
  {034502} (\bibinfo {year} {2014}{\natexlab{a}})},\ \Eprint
  {http://arxiv.org/abs/arXiv:1403.1965} {arXiv:1403.1965 [hep-lat]}
  \BibitemShut {NoStop}%
\bibitem [{\citenamefont {Shanahan}\ \emph
  {et~al.}(2014{\natexlab{b}})\citenamefont {Shanahan} \emph
  {et~al.}}]{Shanahan:2014uka}%
  \BibitemOpen
  \bibfield  {author} {\bibinfo {author} {\bibfnamefont {P.~E.}\ \bibnamefont
  {Shanahan}} \emph {et~al.},\ }\href@noop {} {\bibfield  {journal} {\bibinfo
  {journal} {Phys. Rev.}\ }\textbf {\bibinfo {volume} {D89}},\ \bibinfo {pages}
  {074511} (\bibinfo {year} {2014}{\natexlab{b}})},\ \Eprint
  {http://arxiv.org/abs/1401.5862} {arXiv:1401.5862 [hep-lat]} \BibitemShut
  {NoStop}%
\bibitem [{\citenamefont {Horsley}\ \emph {et~al.}(2013)\citenamefont {Horsley}
  \emph {et~al.}}]{Horsley:2013wqa}%
  \BibitemOpen
  \bibfield  {author} {\bibinfo {author} {\bibfnamefont {R.}~\bibnamefont
  {Horsley}} \emph {et~al.},\ }\href@noop {} {\ \textbf {\bibinfo {volume}
  {PoS}},\ \bibinfo {pages} {249} (\bibinfo {year} {LATTICE 2013})},\ \Eprint
  {http://arxiv.org/abs/1311.5010} {arXiv:1311.5010 [hep-lat]} \BibitemShut
  {NoStop}%
\bibitem [{\citenamefont {Bietenholz}\ \emph {et~al.}(2011)\citenamefont
  {Bietenholz} \emph {et~al.}}]{Bietenholz:2011qq}%
  \BibitemOpen
  \bibfield  {author} {\bibinfo {author} {\bibfnamefont {W.}~\bibnamefont
  {Bietenholz}} \emph {et~al.},\ }\href {\doibase 10.1103/PhysRevD.84.054509}
  {\bibfield  {journal} {\bibinfo  {journal} {Phys. Rev.}\ }\textbf {\bibinfo
  {volume} {D84}},\ \bibinfo {pages} {054509} (\bibinfo {year} {2011})},\
  \Eprint {http://arxiv.org/abs/1102.5300} {arXiv:1102.5300 [hep-lat]}
  \BibitemShut {NoStop}%
\bibitem [{\citenamefont {Leinweber}(2004)}]{Leinweber:2002qb}%
  \BibitemOpen
  \bibfield  {author} {\bibinfo {author} {\bibfnamefont {D.~B.}\ \bibnamefont
  {Leinweber}},\ }\href {\doibase 10.1103/PhysRevD.69.014005} {\bibfield
  {journal} {\bibinfo  {journal} {Phys. Rev.}\ }\textbf {\bibinfo {volume}
  {D69}},\ \bibinfo {pages} {014005} (\bibinfo {year} {2004})},\ \Eprint
  {http://arxiv.org/abs/hep-lat/0211017} {arXiv:hep-lat/0211017} \BibitemShut
  {NoStop}%
\bibitem [{\citenamefont {Tiburzi}(2009)}]{Tiburzi:2009yd}%
  \BibitemOpen
  \bibfield  {author} {\bibinfo {author} {\bibfnamefont {B.~C.}\ \bibnamefont
  {Tiburzi}},\ }\href {\doibase 10.1103/PhysRevD.79.077501} {\bibfield
  {journal} {\bibinfo  {journal} {Phys. Rev.}\ }\textbf {\bibinfo {volume}
  {D79}},\ \bibinfo {pages} {077501} (\bibinfo {year} {2009})},\ \Eprint
  {http://arxiv.org/abs/0903.0359} {arXiv:0903.0359 [hep-lat]} \BibitemShut
  {NoStop}%
\bibitem [{\citenamefont {Shanahan}\ \emph
  {et~al.}(2013{\natexlab{a}})\citenamefont {Shanahan}, \citenamefont
  {Thomas},\ and\ \citenamefont {Young}}]{Shanahan:2012wa}%
  \BibitemOpen
  \bibfield  {author} {\bibinfo {author} {\bibfnamefont {P.~E.}\ \bibnamefont
  {Shanahan}}, \bibinfo {author} {\bibfnamefont {A.~W.}\ \bibnamefont
  {Thomas}}, \ and\ \bibinfo {author} {\bibfnamefont {R.~D.}\ \bibnamefont
  {Young}},\ }\href {\doibase 10.1016/j.physletb.2012.11.072} {\bibfield
  {journal} {\bibinfo  {journal} {Phys. Lett.}\ }\textbf {\bibinfo {volume}
  {B718}},\ \bibinfo {pages} {1148} (\bibinfo {year} {2013}{\natexlab{a}})},\
  \Eprint {http://arxiv.org/abs/1209.1892} {arXiv:1209.1892 [nucl-th]}
  \BibitemShut {NoStop}%
\bibitem [{\citenamefont {Shanahan}\ \emph {et~al.}(2012)\citenamefont
  {Shanahan}, \citenamefont {Thomas},\ and\ \citenamefont
  {Young}}]{Shanahan:2013cd}%
  \BibitemOpen
  \bibfield  {author} {\bibinfo {author} {\bibfnamefont {P.~E.}\ \bibnamefont
  {Shanahan}}, \bibinfo {author} {\bibfnamefont {A.~W.}\ \bibnamefont
  {Thomas}}, \ and\ \bibinfo {author} {\bibfnamefont {R.~D.}\ \bibnamefont
  {Young}},\ }\href@noop {} {\bibfield  {journal} {\bibinfo  {journal} {PoS}\
  }\textbf {\bibinfo {volume} {LATTICE2012}},\ \bibinfo {pages} {165} (\bibinfo
  {year} {2012})},\ \Eprint {http://arxiv.org/abs/1301.3231} {arXiv:1301.3231
  [hep-lat]} \BibitemShut {NoStop}%
\bibitem [{\citenamefont {Shanahan}\ \emph
  {et~al.}(2013{\natexlab{b}})\citenamefont {Shanahan}, \citenamefont
  {Thomas},\ and\ \citenamefont {Young}}]{Shanahan:2013vla}%
  \BibitemOpen
  \bibfield  {author} {\bibinfo {author} {\bibfnamefont {P.~E.}\ \bibnamefont
  {Shanahan}}, \bibinfo {author} {\bibfnamefont {A.~W.}\ \bibnamefont
  {Thomas}}, \ and\ \bibinfo {author} {\bibfnamefont {R.~D.}\ \bibnamefont
  {Young}},\ }\href {\doibase 10.1103/PhysRevD.87.094515} {\bibfield  {journal}
  {\bibinfo  {journal} {Phys. Rev.}\ }\textbf {\bibinfo {volume} {D87}},\
  \bibinfo {pages} {094515} (\bibinfo {year} {2013}{\natexlab{b}})},\ \Eprint
  {http://arxiv.org/abs/1303.4806} {arXiv:1303.4806 [nucl-th]} \BibitemShut
  {NoStop}%
\bibitem [{\citenamefont {Leutwyler}(1996)}]{Leutwyler:1996qg}%
  \BibitemOpen
  \bibfield  {author} {\bibinfo {author} {\bibfnamefont {H.}~\bibnamefont
  {Leutwyler}},\ }\href {\doibase 10.1016/0370-2693(96)00386-3} {\bibfield
  {journal} {\bibinfo  {journal} {Phys. Lett.}\ }\textbf {\bibinfo {volume}
  {B378}},\ \bibinfo {pages} {313} (\bibinfo {year} {1996})},\ \Eprint
  {http://arxiv.org/abs/hep-ph/9602366} {arXiv:hep-ph/9602366} \BibitemShut
  {NoStop}%
\bibitem [{\citenamefont {Aoki}\ \emph {et~al.}(2014)\citenamefont {Aoki} \emph
  {et~al.}}]{FLAG}%
  \BibitemOpen
  \bibfield  {author} {\bibinfo {author} {\bibfnamefont {S.}~\bibnamefont
  {Aoki}} \emph {et~al.},\ }\href
  {http://dx.doi.org/10.1140/epjc/s10052-011-1695-1} {\bibfield  {journal}
  {\bibinfo  {journal} {Eur. Phys. J.}\ }\textbf {\bibinfo {volume} {C74}},\
  \bibinfo {pages} {2890} (\bibinfo {year} {2014})},\ \Eprint
  {http://arxiv.org/abs/arXiv:1310.8555 [hep-lat]} {arXiv:1310.8555 [hep-lat]}
  \BibitemShut {NoStop}%
\bibitem [{\citenamefont {Abdel-Rehim}\ \emph {et~al.}(2014)\citenamefont
  {Abdel-Rehim} \emph {et~al.}}]{Abdel-Rehim:2013wlz}%
  \BibitemOpen
  \bibfield  {author} {\bibinfo {author} {\bibfnamefont {A.}~\bibnamefont
  {Abdel-Rehim}} \emph {et~al.},\ }\href@noop {} {\bibfield  {journal}
  {\bibinfo  {journal} {Phys. Rev.}\ }\textbf {\bibinfo {volume} {D89}}
  (\bibinfo {year} {2014})},\ \Eprint {http://arxiv.org/abs/1310.6339}
  {arXiv:1310.6339 [hep-lat]} \BibitemShut {NoStop}%
\bibitem [{\citenamefont {Wang}\ \emph {et~al.}(2014)\citenamefont {Wang},
  \citenamefont {Leinweber},\ and\ \citenamefont {Thomas}}]{Wang:2014nhf}%
  \BibitemOpen
  \bibfield  {author} {\bibinfo {author} {\bibfnamefont {P.}~\bibnamefont
  {Wang}}, \bibinfo {author} {\bibfnamefont {D.~B.}\ \bibnamefont {Leinweber}},
  \ and\ \bibinfo {author} {\bibfnamefont {A.~W.}\ \bibnamefont {Thomas}},\
  }\href {\doibase 10.1103/PhysRevD.89.033008} {\bibfield  {journal} {\bibinfo
  {journal} {Phys. Rev.}\ }\textbf {\bibinfo {volume} {D89}},\ \bibinfo {pages}
  {033008} (\bibinfo {year} {2014})}\BibitemShut {NoStop}%
\bibitem [{\citenamefont {Nakamura}\ and\ \citenamefont
  {St\"uben}(2010)}]{Nakamura:2010qh}%
  \BibitemOpen
  \bibfield  {author} {\bibinfo {author} {\bibfnamefont {Y.}~\bibnamefont
  {Nakamura}}\ and\ \bibinfo {author} {\bibfnamefont {H.}~\bibnamefont
  {St\"uben}},\ }\href@noop {} {\bibfield  {journal} {\bibinfo  {journal}
  {PoS}\ }\textbf {\bibinfo {volume} {LATTICE2010}},\ \bibinfo {pages} {040}
  (\bibinfo {year} {2010})},\ \Eprint {http://arxiv.org/abs/1011.0199}
  {arXiv:1011.0199 [hep-lat]} \BibitemShut {NoStop}%
\bibitem [{\citenamefont {Boyle}(2009)}]{Boyle:2009vp}%
  \BibitemOpen
  \bibfield  {author} {\bibinfo {author} {\bibfnamefont {P.~A.}\ \bibnamefont
  {Boyle}},\ }\href {\doibase 10.1016/j.cpc.2009.08.010} {\bibfield  {journal}
  {\bibinfo  {journal} {Comput. Phys. Commun.}\ }\textbf {\bibinfo {volume}
  {180}},\ \bibinfo {pages} {2739} (\bibinfo {year} {2009})}\BibitemShut
  {NoStop}%
\bibitem [{\citenamefont {Edwards}\ and\ \citenamefont
  {Joo}(2005)}]{Edwards:2004sx}%
  \BibitemOpen
  \bibfield  {author} {\bibinfo {author} {\bibfnamefont {R.~G.}\ \bibnamefont
  {Edwards}}\ and\ \bibinfo {author} {\bibfnamefont {B.}~\bibnamefont {Joo}}
  (\bibinfo {collaboration} {SciDAC Collaboration, LHPC Collaboration, UKQCD
  Collaboration}),\ }\href {\doibase 10.1016/j.nuclphysbps.2004.11.254}
  {\bibfield  {journal} {\bibinfo  {journal} {Nucl. Phys. Proc. Suppl.}\
  }\textbf {\bibinfo {volume} {140}},\ \bibinfo {pages} {832} (\bibinfo {year}
  {2005})},\ \Eprint {http://arxiv.org/abs/hep-lat/0409003}
  {arXiv:hep-lat/0409003} \BibitemShut {NoStop}%
\end{thebibliography}%

\end{document}